\def\ii{\'{\i}}
\def\beg{\begin{equation}}
\def\bega{\begin{eqnarray}}
\def\begal{\begin{eqnarray}\label}
\def\begl{\begin{equation}\label}
\def\fim{\end{equation}}
\def\fima{\end{eqnarray}}

\documentstyle[aps,preprint]{revtex} 
\tightenlines 

\begin{document}

\title{Phase diagram of a probabilistic cellular automaton with three-site
interactions}

\author{A.P.F. Atman  \thanks{atman@fisica.ufmg.br},
        Ronald Dickman \thanks{dickman@fisica.ufmg.br}  
and     J.G. Moreira  \thanks{jmoreira@fisica.ufmg.br}
        } 
        
\address{Departamento de F\ii sica, Instituto de
        Ci\^encias Exatas, Universidade Federal de Minas Gerais, C. P. 702\\
        30123-970, Belo Horizonte, MG - Brazil}

\date{\today}
\maketitle

\begin{abstract}
We study a (1+1) dimensional probabilistic cellular automaton that is
closely related to the Domany-Kinzel  (DKCA), but in which the update of 
a given site depends on the state of {\it three} sites at the previous time 
step. Thus, compared with the DKCA, there is an additional parameter, $p_3$, 
representing the probability for a site to be active at time $t$, given that 
its nearest neighbors and itself were active at time $t-1$. 
We study phase transitions and 
critical behavior for the activity {\it and} for damage spreading,
using one- and two-site mean-field approximations, and simulations,
for $p_3=0$ and $p_3=1$. We find evidence for a 
line of tricritical points in the ($p_1,~p_2,~p_3$) parameter space, 
obtained using a mean-field approximation at pair level.
To construct the phase diagram in simulations
we employ the growth-exponent method in an interface representation. 
For $p_3 =0$, the phase diagram is similar to the DKCA, but the damage 
spreading transition exhibits a reentrant phase.
For $p_3=1$, the growth-exponent method reproduces the
two absorbing states, first and second-order phase transitions, bicritical point,
and damage spreading transition recently identified by
Bagnoli {\it et al.} [Phys. Rev. E{\bf 63}, 046116 (2001)].
\vspace{1pc}

\vspace{1pc}
\end{abstract}

\section{Introduction}

Probabilistic cellular automata (PCA) are widely used to model
systems with local interactions in physics, chemistry, biology and social 
sciences \cite{wolfram,farmer,mann,guto,boccara}. 
Despite their simplicity, these models
exhibit complex behavior and are used to investigate fundamental problems
in statistical mechanics, such as spin models 
\cite{bagstat,georges} and nonequilibrium phenomena \cite{dk,bagnoli}. In particular, 
the problem of phase transitions in the presence of absorbing states has attracted
increasing interest in recent years; PCA play a major role in 
these studies \cite{bhatta,atman2,mario,kwon}. The PCA 
introduced by Domany and Kinzel \cite{dk} is, along with 
the contact process \cite{cp,mundick}, one of the simplest 
models exhibiting an absorbing-state phase transition.

The one-dimensional Domany-Kinzel stochastic cellular automaton (DKCA) is a 
completely discrete sys\-tem - temporally, spatially and in its state space - 
which attracts interest as a particle
system affording a test of ideas on scaling in nonequilibrium critical
phenomena \cite{kinzel}. 
The DKCA has a unique absorbing (``vacuum") state; its phase diagram presents a 
critical line separating this absorbing phase from an active one.
Continuous phase transitions to an absorbing state are conjectured to belong generically 
to the directed percolation (DP) universality class \cite{grass1}. 
In addition to the active-absorbing transition, Martins {\it et al.} \cite{martins} 
found a damage spreading (DS) transition 
separating the active phase into nonchaotic and chaotic phases. 
There is numerical evidence that the critical  
behavior along this transition line also belongs to the DP class, as 
expected on the basis of universality \cite{grass2}. 

Recently, Bagnoli {\it et al} \cite{bagnoli} introduced a model that can be 
considered a natural extension of the DKCA: a one-dimensional PCA 
in which the update of a given site depends on the state of its nearest neighbors 
and itself, at the preceding time step. 
(We shall refer to this model as the BPCA.)
Thus, compared with the DKCA, there is an additional parameter, $p_3$, 
representing the probability for a site to be active at time $t$, given that 
all three sites were active at time $t-1$. Bagnoli {\it et al.} studied 
$p_3=1$, in which case the model presents two absorbing states: the empty one and the 
completely occupied configuration. As in the DKCA, the density is the order 
parameter.  These authors used the mean field approximation (at site level), 
simulations, and field-theoretic arguments to study the model. They found a 
rich phase diagram, with first- and second-order phase transitions, a 
bicritical point, and a damage-spreading transition. Except for the line 
$p_2=1$ in the DKCA, this is the simplest PCA
that exhibits a discontinuous phase transition \cite{bagnoli}.

In this work we extend the analysis of the BPCA
considering two cases: the previously studied $p_3=1$, which corresponds to a 
ferromagnetic-like model, and $p_3=0$, representing a 
{\it game-of-life}-like model \cite{conway,baglife}. We extend the mean-field 
analysis to the pair level, and use simulations to construct the
phase diagram. In simulations, we apply the growth-exponent method 
\cite{atman} to identify transitions.

This paper is structured as follows: in Section II we define the model and its 
interface representation; the site and pair mean-field approximations are 
discussed in Section III. Simulation results are presented
in Section IV. We summarize our findings in Section V.

\section{Model}

The one-dimensional PCA with three-site neighborhood (BPCA) was proposed by 
Bagnoli {\it et al.} \cite{bagnoli}.  It consists of a ring of $L$ sites 
($i=1,2,...,L$), with periodic boundaries, in which each site $i$ has two possible 
states, conveniently denoted by $\sigma_i = 0,1$. The state of the system at 
time $t$ is given
by the set \{$\sigma_{i}(t)$\}. In contrast to the deterministic CA studied by 
Wolfram \cite{wolfram}, the present model is a discrete time Markov process: 
the rules for updating the system are given by transition probabilities.
In particular, state of site $i$ at time $t\!+\!1$ 
depends on $\sigma_{i\!-\!1}(t)$, $\sigma_i(t)$ and $\sigma_{i\!+\!1}(t)$, via 
the transition probability 
$P( \sigma_{i}(t\!+\!1) | \sigma_{i\!-\!1}(t),\sigma_i(t),\sigma_{i\!+\!1}(t) )$.
The latter is of
totalistic form, i.e., the dependence is through 
$S_i(t) = \sigma_{i\!-\!1}(t)+\sigma_i(t)+\sigma_{i\!+\!1}(t)$.
Since $S(t)\!=\!0$ implies $\sigma_{i}(t\!+\!1)=0$ with probability 1,
there remain three free parameters for defining the transition probability.
Specifically: \\
$P(1|0,0,1) = P(1|0,1,0) = P(1|1,0,0) = p_1~,$ \\
$P(1|0,1,1) = P(1|1,0,1) = P(1|1,1,0) = p_2~,$ \\
$P(1|1,1,1) = p_3~.$ \\
Evidently, 
$P(0| \sigma_{i-1},\sigma,\sigma_{i+1} ) = 1 - P(1| \sigma_{i-
1},\sigma_i,\sigma_{i+1})$.

Depending on the values of ($p_1,p_2,p_3$), the asymptotic 
($t \rightarrow \infty$) state of the system is either in an absorbing 
phase ({\it phase 0}, with all sites in state 0, or {\it phase 1}, with all sites 
in state 1), or in the active phase, in which the stationary density $\rho$ of 
sites in state 1 takes a value different from zero or one.
Complete determination of the phase diagram in the three-dimensional 
parameter space is a rather difficult open problem. 
In this work we focus on two cases: $p_3=1$ and $p_3=0$. 
In the first case, the model possesses the two absorbing phases cited 
above, as well as an active phase and a chaotic region (associated with
damage spreading).  For $p_3=0$,
the phase 1 is no longer absorbing (though the phase 0 of course remains so), 
and there is again an active phase; the chaotic region is reentrant.
$p_3=0$ describes a situation in which ``crowding" of individuals leads to their
destruction, similar to Conway's Game of Life model \cite{conway,baglife}, 
while $p_3=1$ corresponds to a ferromagnetic-like model. 

The {\it absorbing/active} transitions are continuous phase transitions, 
characterized by critical exponents which belong to the DP universality class.
The {\it phase 0/phase 1} 
transition is discontinuous \cite{bagnoli}, 
and the exponents are those of {\it compact} directed percolation.
The DS transitions are also in the DP class, consistent with  
Grassberger's prediction \cite{grass2}. The termination of two 
critical lines at a line of discontinuous transitions marks a 
{\it bicritical} point, as has been found in the BPCA for $p_3=1$.
For $p_3 \!<\! 1$, the phase 1 is no longer absorbing, so that
one of the phase boundaries (i.e., between the active and phase 1-absorbing
phases) is no longer present.  We find (using the two-site mean-field
approximation, discussed in Sec. IIIB), that the bicritical point is actually
one terminus of a line of {\it tricritical} points: for each fixed $p_3$ in
the range $1 > p_3 > p_3^t \simeq 1/3$, the absorbing-active transition
is discontinuous for $(p_1,p_2) < (p_1^t,p_2^t)$ and continuous for 
larger values, $(p_1,p_2) > (p_1^t,p_2^t)$, where $(p_1^t,p_2^t)$ is the 
tricritical point. For $p_3 \le p_3^t$, the absorbing/active
transition is always continuous.  

\subsection{Surface Representation}

The mapping of dynamical systems to a surface-growth representation is an 
interesting problem, since in many cases the resulting scaling properties are 
unknown {\it a priori}. Integration of the local activity (with respect to time) 
is the most natural procedure.  
The present method employs the interface
representation proposed by de Salles {\it et al.} 
\cite{sales1}. The procedure consists in transforming the 
spatiotemporal patterns 
generated by the PCA to a solid-on-solid (SOS) particle deposition.
The surface-growth process is attended by 
kinetic roughening; the associated critical exponents can be measured \cite{atman2} 
following the scaling concepts developed by Family and Vics\'ek \cite{family}. 
Atman and Moreira \cite{atman} demonstrated that the growth exponent $\beta_w$ 
exhibits a cusp at criticality, and is very useful for detecting phase 
transitions. 

Height variables are defined by summing the variables 
$\sigma_{i}(\tau)$ over the first $t$ time 
steps:
\beg\label{soma}
h_{i}(t) \equiv \sum^{t}_{\tau=0} \sigma_{i}(\tau)~~.
\fim
In this way we generate a growth process, with correlations 
embodied in the roughness $w(L,t)$ \cite{barab},
defined by
\beg\label{rugos}
w^{2}(L,t) = \frac{1}{L} \left< \sum^{L}_{i=1} \left( h_{i}(t) - \overline{h}(t) 
\right)^{2} \right>~~,
\fim
\noindent where $\overline{h}(t)$ is the mean value of $h_{i}(t)$ at time $t$,
and $<\ldots>$ denotes an average over realizations.

For surface growth models \cite{barab}, 
we expect  $w(L,t)$ to follow the scaling form \cite{family},
\beg\label{scal}
w(L,t) \sim L^{\alpha} f \left( \frac{t}{L^{z}} \right)~~,
\fim
\noindent where $f(u)$ is a universal scaling function, $\alpha$ is the 
roughness exponent, $z = \alpha/ \beta_w$ the dynamic exponent and 
$\beta_w$ the growth exponent. The function $f(u) = constant$ for large $u$,
while $f(u) \sim u^{\beta_w}$ for small $u$ 
($t \ll L^{z}$).
At short times, therefore, we expect $w(t) \sim t^{\beta_w}$.
It is possible to  
measure $\beta_w$ from the slope of a $\log - \log$ plot of 
$w(L,t)$ versus $t$. 
In the active phase, the roughness does not saturate, growing instead as  
$w(L,t) \sim t^{1/2}$, corresponding to uncorrelated growth \cite{atman}, 
since the correlation length is finite, away from the critical point.

In previous work, Atman and Moreira \cite{atman} showed that $\beta_w$ 
attains a maximum at the phase transition, and measured its value along 
the transition line of the DKCA. DP values for others critical
exponents in the surface representation of DKCA are verified in 
Ref. \ref{atman2}.

\subsection{Damage Spreading}

Martins {\it et al.} \cite{martins} used the damage-spreading technique 
to show that the active phase of the DKCA consists of two phases, chaotic and 
nonchaotic. The order parameter of this transition is the difference between 
two replicas initialized with different configurations, but subject to
the same sequence of random events during the subsequent evolution.
In practice, we let the system evolve until it attains a stationary state, 
and then copy the configuration, introducing some alterations (``damage"). 
The two replicas, one with 
state $\sigma_{i}(t)$ and the other with state $\varrho_{i}(t)$, then 
evolve using the same sequence of random numbers, and the difference 
between them,  
\[
\Gamma_{i}(t) = | \sigma_{i}(t) - \varrho_{i}(t) |~,
\]
is monitored. The fraction of sites in the two replicas with 
$\sigma_i \neq \varrho_i$ defines their Hamming distance: 
\begl{def:D_H}
D_H (t) = \frac{1}{L}\sum_i \Gamma_i (t)~.
\fim
The stationary Hamming distance is null in the non-chaotic phase and 
positive in the chaotic phase. 

To study the chaotic/non-chaotic boundary, we use a slightly different method, 
in which the {\it difference} between the replicas is used to generate 
a {\it surface growth process}, as described above.  In this case,
\beg\label{acc2}
h_{i}(t) = \sum^{t}_{\tau=0} \Gamma_{i}(\tau)~.
\fim
Thus, the profile generated by the difference between the replicas behaves 
just as the profiles at the phase 0/active boundary: the roughness 
reaches a stationary value in the non-chaotic phase and grows indefinitely in 
the chaotic phase, with a cusp in the $\beta_w$ value at the transition.

Since the system has already relaxed to the stationary state when
we initiate the damage experiment, creating the damage by randomly
altering sites in one copy is likely to perturb the particle density
and correlation functions away from their stationary values.  This
in turn would introduce an undesirable assymetry between the replicas,
since the dynamics of damage spreading would be mixed with that of 
relaxation back to the stationary state (in the copy but not in the
original).  To avoid such complications,
we generate the damage by {\it rotating} the copy by 
180$^0$ with respect to the original, 
with no further modifications,
that is, $\varrho (i, t_0) = 
\sigma(i + L/2, t_0)$, subject to periodic boundary conditions. 
This represents a large initial damage (a Hamming distance of
$\simeq 2 \rho(1 \!-\! \rho)$, with $\rho$ the stationary 
particle density), which is statistically uniform over the system.

\section{Mean-Field Theory}

\subsection{One-site approximation}

To begin, we present the mean-field approximation at the site level, 
for $p_3=1$ and $p_3=0$, and construct the phase diagram from the 
equations obtained in this approximation.
The BPCA is a Markov process in which all sites are updated simultaneously. 
The configuration $\{ \sigma \}$ is a set of stochastic variables with 
probability distribution at time $t$ given by $P_t(\sigma)$. The evolution 
of the latter is governed by
\beg\label{eq:Pt}
P_{t+1} = \sum_{\sigma'} \omega(\sigma|\sigma') P_{t} (\sigma')~,
\fim
where $\omega(\sigma|\sigma')$ denotes the probability of the transition 
$\sigma' \to \sigma$, with the properties $\omega(\sigma|\sigma') \ge 0$ and
$\sum_{\sigma} \omega(\sigma|\sigma') =1$. The transition probability for the
BPCA is a product of factors associated with each site:
\beg\label{eq:wi}
\omega(\sigma|\sigma') = \prod_{i=1}^L w_i (\sigma_i|\sigma')~,
\fim
where $w_i(\sigma_i|\sigma') \ge 0$ is the conditional probability for  
site $i$ to be active at time $t+1$ given the configuration $\sigma'$ at 
the preceding step.
The probabilities $w_i$ are translationally invariant and in fact depend 
only on the states $\sigma_{i-1}$, $\sigma_i$ and $\sigma_{i+1}$ at the 
previous step:
\beg\label{eq:w3s}
w_i(\sigma_i|\sigma') = w_{3s}(\sigma_i|\sigma'_{i-1}, \sigma'_i, \sigma'_{i+1})~.
\fim
 We list the $w_{3s}$ in Table I.
 
Of interest are the $n$-site marginal probabilities. The evolution of
the one-site distribution $P_t(\sigma_i)$, is given by
\beg\label{seis} 
P_{t+1}(\sigma_i) = \sum_{\sigma'_{i-1}}\sum_{\sigma'_i}\sum_{\sigma'_{i+1}}
w_{3s}(\sigma_i|\sigma'_{i-1}, \sigma'_i, \sigma'_{i+1}) 
 P_t(\sigma'_{i-1}, \sigma'_i, \sigma'_{i+1})~,
\fim 
where $ P_t(\sigma'_{i-1}, \sigma'_i, \sigma'_{i+1})$ is the marginal 
distribution for a set of three nearest-neighbor sites.  The evolution 
of the two-site distribution is given by:
\bega 
\nonumber
P_{t+1}(\sigma_i,\sigma_{i+1})  &=&  
\sum_{\sigma'_{i-1}}\sum_{\sigma'_i}\sum_{\sigma'_{i+1}} \sum_{\sigma'_{i+2}} 
w_{3s}(\sigma_i|\sigma'_{i-1}, \sigma'_{i}, \sigma'_{i+1}) \\
\label{dez}
&\times& w_{3s}(\sigma_{i+1}|\sigma'_{i}, \sigma'_{i+1}, \sigma'_{i+2}) 
P_t(\sigma'_{i-1}, \sigma'_{i}, \sigma'_{i+1},\sigma'_{i+2})~. 
\fima 
Evidently we have an infinite hierarchy of equations.  In the $n$-site
approximation the hierarchy is truncated by estimating the
$(n\!+\!1)$-site (and higher) probabilities on the basis of those for 
$n$ sites.

The simplest case is the one-site approximation, in which 
$P_t(\sigma'_{i-1}, \sigma'_{i},\sigma'_{i+1})$ 
is factored so:   
$P_t(\sigma'_{i-1}, \sigma'_{i},\sigma'_{i+1}) = P_t(\sigma'_{i-1}) P_t(\sigma'_{i}) P_t(\sigma'_{i+1})$.
This yields the recurrence relation,  
\beg\label{xum} 
\rho_{t+1} = p_3 \rho_t^3 + 3p_2 \rho_t^2(1-\rho_t) +3 p_1 \rho_t(1-\rho_t)^2~,
\fim 
where $\rho_t \equiv P_t(1)$ is the density of active sites
(i.e., the order parameter).

Depending on the value of ($p_1,~p_2,~p_3$), Eq. (\ref{xum}) admits different 
stationary solutions, corresponding to the possible BPCA phases discussed
above: {\it phase 0} ($\rho = 0$), {\it phase 1} ($\rho = 1$) and {\it active} 
($0 < \rho < 1$). In order to verify the stability of the stationary 
solutions, we consider a small perturbation in the stationary value $\rho^*$,
$\rho_t = \rho^* + \Delta \rho_t$. Applying this variable change in the 
Eq. (\ref{xum}), we obtain for the mean-field approximation at the site level,
\bega
\nonumber
\Delta \rho_{t+1} & = &  \Delta \rho_t [ 3{\rho^*}^2 (p_3+ 3p_1 -3p_2) + 6 \rho^* 
(p_2 -2p_1) + 3p_1] + \\
\nonumber
& & (\Delta \rho_t)^2 [ 3\rho^* (p_3+ 3p_1 -3p_2) + 3 (p_2 -2p_1) ] + \\
\label{Dro}
& & (\Delta \rho_t)^3 [ p_3 + 3p_1 - 3p_2 ] ~.
\fima
We can write Eq. (\ref{Dro}) in a simplified manner, 
\[
\Delta \rho_{t+1} = a(\rho^*) \Delta \rho_t + b(\rho^*) \Delta^2 \rho_t + 
c \Delta^3 \rho_t ~,
\]
where the coefficients $a(\rho^*)$, $b(\rho^*)$ and $c$ can be associated
with the stability of the solutions. 

Considering the solution, $\rho^* = 0$, the stability condition is 
$a(\rho^*) \! < \! 1$, which implies $p_1 \! < \! 1/3$. In the case 
$a(\rho^*)\! =\! 1$, 
$(p_1 \! = \! 1/3)$, the solution will be stable only if $b(\rho^*)\! < \! 0$ and 
$c \! < \! 0$;
the first condition implies $p_3 \! < \! 2/3$. For $a(\rho^*)\! = \! 1$ and 
$b(\rho^*)\! = \! 0$, the stability
the condition $c \! < \! 0$ leads to $p_3 \! < \! 1$. Thus, the solution 
$\rho^* \! = \! 0$
is always stable for $p_1 \! < \! 1/3$, $p_2 \! < \! 2/3$ and $p_3 \! < \! 1$. 
The point (1/3,~2/3,~1) corresponds to a tricritical point in this aproximation.

Considering the solution, $\rho^* \! = \! 1$, the stability condition 
$a(\rho^*) \! < \! 1$ implies $p_3 -p_2 \! < \! 1/3$. In case 
$a(\rho^*) \! = \! 1$, $(p_3 \! = \! p_2 + 1/3)$, the condition 
$b(\rho^*) \! < \! 0$ yields  $p_2 - p_1 \! > \! 1/3$. For 
$a(\rho^*) \! = \! 1$ and $b(\rho^*) \! = \! 0$, the stability
condition $c \! < \! 0$ implies $p_3 \! < \! 1$. So, the point 
(1/3,~2/3,~1) is also a tricritical point for the solution $\rho^* \! = \!1$! 
Thus, the point (1/3,~2/3,~1) corresponds to a {\it bicritical} point, as 
Bagnoli {\it et al.} have
already shown. In fact, the solution $\rho^*=1$ is absorbing only for 
$p_3 \! = \!1$ (since for $p_3 \! < \! 1$, the dynamics of updating destroys
the phase 1), and this solution is stable for $p_2 \! > \! 2/3$.

For $0 < \rho^* < 1$, the stability condition $a(\rho^*) < 1$ implies
a inequality of second degree in terms of $\rho^*$. Considering $a=1$, we
can solve the corresponding equation, and obtain 
\[
\rho^* = \frac{(2p_1 - p_2) \pm \sqrt{(p_2 - 2p_1)^2 - (p_3 +3p_1 - 3p_2)
(3p_1 -1)}}{p_3 - 3p_1 -3p_2} ~.
\]
Note that $\rho^*$ vanishes on the line $(1/3,2/3,p_3)$. It is easy to see 
that the positive solution, $\rho^*_+$,  is valid for any 
($p_1,~p_2,~p_3$), but the negative solution, $\rho^*_-$, is valid only 
for $p_2 < 2 p_1$. Considering the plane $p_1=1/3$, the transition 
line for $p_2 > 2/3$ coincides with the vanishing of the square root, since
$\rho^*$ must be real. This implies a discontinuous 
transition for $p_2 > 2/3$, and the line $(1/3,2/3,p_3)$ corresponds to the
tricritical line in ($p_1,~p_2,~p_3$) space, in the site approximation.

In the simulations we consider two cases: $p_3 =1$ and $p_3=0$. Considering
the stability analysis above, we can summarize the 
phase diagram in these two cases in the following way:
for $p_3=1$, Eq. (\ref{Dro}) can be written as,
\beg\label{eq:xp31} 
\rho \left( (3p_1 - 3p_2 +1) \rho^2 + (3p_2 - 6p_1) \rho + 3p_1 - 1 \right )= 0~. 
\fim 
The three solutions of this equation are:
\begin{itemize}
\item {$\rho = 0$ - Phase 0, stable for $p_1 < 1/3$ and $p_2 < 2/3$;}
\item {$\rho = 1$ - Phase 1, stable for $p_2 > 2/3$ and $p_1 > 1/3$;}
\item {Active phase, for $p_2 < 2/3$ and $p_1 > 1/3$, where the 
stationary density is given by:
\beg\label{eq:xacp31}
\rho = \frac{3p_1 - 1}{3p_1 -3p_2 + 1}~.
\fim
}
\end{itemize}
For $p_2 > 2/3$ and $p_1 < 1/3$, we have a discontinuous transition line 
separating the phase 0/phase 1 (both stable) at $p_2 = 1-p_1$. 

In case $p_3 = 0$, we have:
\begl{eq:xp30}
\rho \left( (3p_1 - 3p_2) \rho^2 + (3p_2 - 6p_1) \rho + (3p_1 - 1) \right) = 0~.
\fim 
For this equation there are only two distinct phases: 
\begin{itemize}
\item {$\rho = 0$ - Frozen phase, stable for $p_1 < 1/3$, $p_2 \le 2/3$;}
\item {Active phase, valid for $p_1 > 1/3$ and $p_1 \neq p_2$, 
where the stationary density is given by: 
\begl{eq:xacp30}
\rho = \frac{6 p_1 - 3p_2 \pm \sqrt{9 p_2^2 -12 p_2 + 12 p_1}}{6 p_1 - 6 p_2}~;
\fim
the negative root is valid for $p_2 < 2p_1$, while the positive root is 
valid for $p_2 > 2p_1$.
}
\end{itemize}
For $p_2 > 2/3$, the solution of eq. (\ref{eq:xacp30}) is
either complex or strictly $>0$, implying a {\it discontinuous} 
transition from the phase 0 to the active phase, as we antecipate. 

The phase diagram in the one-site approximation, for $p_3=1$, is shown in 
Fig. 1; that for $p_3=0$ is shown in Fig. 2. For $p_3 =1$, the phase diagram
is as expected \cite{bagnoli}. For $p_3=0$, we expect a single 
absorbing state, but the behavior for $p_2 > 2/3$, 
where the transition line is discontinuous is not expected on the basis of
simulations. (At the transition the r.h.s. of Eq. (\ref{eq:xacp30}) changes
from a complex to real, nonzero value.) 

\subsection{Pair approximation}

At the pair level, the probability 
$P_t(\sigma'_{i-1}, \sigma'_{i}, \sigma'_{i+1},\sigma'_{i+2})$ is factored in the 
following way
\begl{fac:pair}
P_t(\sigma'_{i-1}, \sigma'_{i}, \sigma'_{i+1},\sigma'_{i+2}) = 
\frac{P_t(\sigma'_{i-1}, \sigma'_{i}) P_t(\sigma'_{i}, \sigma'_{i+1}) P_t(\sigma'_{i+1}, \sigma'_{i+2})}
{P_t(\sigma'_{i}) P_t(\sigma'_{i+1})}~.
\fim
Calling $z \equiv P(1,1)$, and noting that 
$P(1) = P(1,0)+ P(1,1)$, we can write $P(1,0) = P(0,1) = \rho \!-\! z$, and
$P(0,0) \equiv 1 -2\rho +z$.
The recursion relations for the 
density of active sites $\rho$ and for the density of 
active pairs $z$ are:

\bega
\nonumber
\rho_{t+1} & = & p_3 \frac{z_t^2}{\rho_t} + p_2 (\rho_t-z_t)
\left(\frac{2 z_t }{\rho_t} + \frac{(\rho_t-z_t)}{1-\rho_t} \right)  \\
\label{eq:xsite}
& + &  p_1  (\rho_t-z_t) \left( \frac{2 (1-2\rho_t+z_t)}{1-\rho_t} + 
\frac{(\rho_t-z_t)}{\rho_t} \right)~,
\fima

\bega
\nonumber
z_{t+1} & = & p_1^2 \left( \frac{(\rho_t-z_t)^2 (1-2\rho_t+z_t) (2-\rho_t)}{\rho_t(1-\rho_t)^2}  \right)\\
\nonumber
& + & 2 p_1 p_2 \left( \frac{(\rho_t-z_t)}{\rho_t(1-\rho_t)} \right) \left( z_t(1-2\rho_t+z_t) + (\rho_t-z_t)^2 \right) \\
\label{eq:zpair}
& + &  p_2^2 \left( \frac{z_t (\rho_t-z_t)^2 (1+\rho_t)}{\rho_t^2 (1-\rho_t)} \right)
+ 2 p_2 p_3 \left(\frac{z_t^2 (\rho_t-z_t)}{\rho_t^2} \right) + p_3^2 \frac{z_t^3}{\rho_t^2}~.
\fima
Iterating these relations numerically until a steady state 
is reached, we construct the phase diagram in the pair approximation. 
Results for $p_3 =1$ and $p_3 = 0$ are shown in Figs. 1 and 2, respectively. 

Using the pair approximation, Eq. (\ref{eq:xsite}) and Eq. (\ref{eq:zpair}), 
we find numerically the critical surface in the ($p_1,p_2,p_3$) parameter 
space and the line of tricritical points, for $p_3<1$, as 
sketched in Fig. 3. In the region $p_3 < 1$, the phase 1 disappears 
and the discontinuous absorbing transition lines meet the continuous transition
lines at the tricritical points, as shown in  
Fig. 3. Each one of the eight vertices in this diagram 
corresponds to a different deterministic rule in the automata studied by 
Wolfram \cite{wolfram}; for example, ($p_1=1,~p_2=0, ~p_3=0$) corresponds to 
the rule 22; ($p_1=1,~p_2=1, ~p_3=0$) to rule 126; etc.

\subsection{Damage Spreading Transition at Site Level}

Bagnoli {\it et al.} \cite{bagnoli} derived a mean-field approximation 
for the DS transition at $p_3 =1$, showing that there is a chaotic region in
the active phase of the BPCA. To obtain the one-site approximation for 
the BPCA at $p_3=0$, we use the approach of Tom\'e 
\cite{tome}; denoting the configurations by 
$\{ \sigma_i \}$ and $\{ \tau_i \}$, 
the Hamming distance is given by
\begl{eq:hamm}
H_{t} =  \langle (\sigma_i - \tau_i)^2 \rangle~,
\fim
where the brakets denote an average over realizations. The evolution of
the joint probability follows
\begl{eq:probhamm}
P_{t+1} (\sigma ; \tau) = \sum_{\sigma, \tau} 
W (\sigma ; \tau \mid \sigma' ; \tau')
P_t (\sigma' ; \tau')~,
\fim
where 
\begl{def:hammrate}
W(\sigma ; \tau \mid \sigma' ; \tau') = \prod_i \varpi(\sigma_i ; \tau_i \mid 
\sigma'_{i-1}, \sigma'_i, \sigma'_{i+1} ; \tau'_{i-1}, \tau'_i, \tau'_{i+1})~,
\fim
is the transition probability for two the two systems 
(subject to the same noise), 
from the state ($\sigma' ; \tau'$) to ($\sigma ; \tau$). 
Using the transition probabilities defined in Table I, we can 
calculate the joint transition probabilities, as shown in Table II.

Now, we can write the equations for the evolution of the order parameter 
associated with the chaotic transition - the Hamming distance. Denoting the 
Hamming distance defined by Eq. (\ref{def:D_H}) as 
$\psi_t \equiv P_t(1;0) = P_t(0;1)$, and using the relation (\ref{eq:probhamm}), 
we have
\[
\psi_{t+1} = < \varpi(1;0 \mid \sigma_{i-1}, \sigma_i, \sigma_{i+1}; \tau_{i+1}, \tau_i, \tau_{i+1})>~;
\]
using the rules of Table II, we can write $\psi_t$ as 
\begal{eq:P10}
P_{t+1}(1;0) & = &  p_3 P_t(1,1,1;0,0,0) + 3 p_2 P_t(1,1,0;0,0,0) + 
 \\ 
\nonumber
& &  + \: 3 (b+c) P_t(1,1,1;1,1,0)+ 3 p_1 P_t(1,0,0;0,0,0) +  \\
\nonumber
& & + \: 3 (b'+ c') P_t(1,1,1;1,0,0) + 9 (b''+ c'') P_t(1,1,0;1,0,0)~.
\fima
Calling $P_t(1) \equiv x_t$, 
we can write $P_t(1;1) = x_t - \psi_t$ and $P_t(0;0) = 1 - x_t -\psi_t$; thus,
using the one-site mean-field approximation 
$P_t(\sigma_{i-1}, \sigma_i, \sigma_{i+1} ; \tau_{i-1}, \tau_i, \tau_{i+1}) =
P_t(\sigma_{i-1}, \sigma_i, \sigma_{i+1}) P_t(\tau_{i-1}, \tau_i, \tau_{i+1})$,
we can write Eq. (\ref{eq:P10}) as
\begal{eq:P10x}
\psi_{t+1} & = & \psi_t [ p_3 \psi_t^2 + 3 p_2 \psi_t (1- x_t -\psi_t) + 3 p_1 (1 -x_t -\psi_t)^2 +
3 (b+c)(x_t + \psi_t)^2 \\
\nonumber
& & \: + 3 (b'+c') \psi_t (x_t - \psi_t) + 9(b''+c'') (x_t - \psi_t)(1 -x_t -\psi_t) ]~.
\fima

Finally, considering the case $p_3 = 0$, we have $(b+c)=p_2$, $(b'+c')=p_1$ and
$(b''+c'')= \mid p_1 - p_2 \mid$; inserting these values in Eq. (\ref{eq:P10x}) we
obtain
\begal{eq:psi0}
\psi_{t+1} & = & 3 \psi_t \{ 3\mid p_1 - p_2\mid \psi_t^2 + 
[(p_2-2p_1) - 3\mid p_1 - p_2 \mid + 3(p_1 - p_2)x_t] \psi_t \\
\nonumber
& & \: + [(p_1 + p_2 -3 \mid p_1 -p_2\mid)x_t^2 + (3\mid p_1 - p_2 \mid -2p_1)x_t + p_1] \}~.
\fima
This equation can be iterated numerically using the stationary values of 
$x_t$  obtained from Eq. (\ref{eq:xp30}); there are three possibilities 
for the joint
solutions of Eq. (\ref{eq:xp30}) and Eq. (\ref{eq:psi0}): 
$x = \psi =0$, corresponding
to the phase 0; $x =0,~\psi \neq 0$, corresponding to the 
active phase; 
and $x,~\psi \neq 0$, corresponding to the chaotic phase. In Fig. 4 
we show 
the stationary solutions for these equations for some values of $p_2$. 
We note that there is a discontinuous DS transition line in this approximation:
for $p_2\! > \!p_t$ ($p_2\! =\! 0.9$ in Fig. 4, for example), $\psi$ is always positive if $x>0$. It implies that the 
DS transition line for $p_2\! >\! p_t$ falls on the discontinuous transition 
analysed in subsection III A.

\section{Simulation Results}

We construct the BPCA phase diagram, for $p_3 =0$ and $p_3=1$, using 
simulations of systems of up to $L=10000$ sites (with periodic boundaries),
applying the growth exponent method \cite{atman} to locate the 
transition lines. The initial condition used in the simulations is 
random, with half the sites occupied. The phase diagrams  
for absorbing-state transitions are shown in Figs. 1 and 2, 
for $p_3=1$ and $p_3=0$, respectively.
As expected, the pair approximation yields a better prediction
than does one-site mean-field theory.
Note that for $p_3 \!=\! 0$, the phase diagram is qualitatively 
the same as for the DKCA \cite{martins,atman}; 
the major difference is that the active phase is quite enlarged 
in the BPCA.

For $p_3 =1$, the pair approximation prediction for the phase boundaries
is qualitatively correct, although the bicritical point 
remains in the same position ($p_1 \!=\! 1/3, p_2 \!=\!2/3$), 
as in the site approximation.
Simulations place the 
bicritical point at (0.460(3),0.540(3)), but the 
phase boundaries are in reasonable agreement with the
pair approximation prediction.
It is important to note that there are only three transitions in this diagram:
the {\it phase 0/phase 1} transition (discontinuous) and the 
{\it phase 0/active} and {\it active/phase 1} transitions (continuous).
All transitions are located using the growth-exponent method, 
confirming that this method is able to detect both continuous and discontinuous 
phase transitions.
 
The phase diagrams for the DS transition are shown in Figs. 5 and 6, for
$p_3=0$ and $p_3=1$ respectively. In the case 
$p_3 = 1$, we confirm the results of Bagnoli {\it et al.}, but some 
comments are in order. In Ref. \cite{bagnoli}, the authors sketched 
several ``damaged domains'' that appear along the active/absorbing 
phase boundary, and attributed them to the divergence
of the relaxation time, or to the fact that small differences in the initial 
configuration can drive the system to a different absorbing state.
As shown in Fig. 7, where we compare the DS transitions obtained
using (1) ``rotation'' damage and (2) random damage in $10 \%$ of the sites, 
these domains are only associated with the absorbing-state transition. We
see two maxima in the $\beta_w \times p_1$ curves in Fig. 7: 
the left maximum, more pronounced, that corresponds to the 
absorbing/active transition, and the right maximum, which actually corresponds 
to the nonchaotic/chaotic transition and belongs to the DP class. 
The left maximum
can yields to an ``apparent'' DS transition, as the `islands' of damage 
commented in the Ref. \cite{bagnoli}.

Thus, the left maximum corresponds to 
the ``damage domains'' of the Ref. \cite{bagnoli}, but in fact it 
is due the absorbing/active transition: in this region, when the replica 
is created (in the stationary state),
it turns out that only a small region of the ring is active, which results in
a constant contribution to the Hamming distance (proportional 
to $2\rho^*$ with rotation damage, and to $2 \rho(t)$, with 
random damage). Thus, with rotation damage, the exponent $\beta_w$ 
approaches 1 because the system is already in the stationary state, 
while with the random damage, $\beta_w \sim 0.84$, due decay of the activity,
recovering the DP value.
This behavior should not be confused with the {\it true} damage transition
that occurs only at the second maximum, and which corresponds to a unique 
phase boundary, as shown in Fig. 6.

In the case $p_3=0$, the simulations confirm the reentrant chaotic transition
predicted by the one-site mean-field approximation. As shown in Fig. 5, 
the DS transition line is concave inward toward the active phase, 
and presents distinct
behaviors for $p_1 > p_2$ and $p_1 < p_2$, as expected. 
We note that the simulation results suggest a sudden cnahge in the orientation
of the active/chaotic phase boundary where the latter crosses the line 
$p_2=p_1$. Such discontinuity of slope, clearly evident in the MF prediction,
may be a consequence of the singular behavior of several transition 
probabilities on $p_1-p_2$ parameters space, as shown in Table II and 
Eq. (\ref{eq:psi0}).

\section{Conclusions}

In this work we apply the growth-exponent method in Monte Carlo simulations, 
and one- and two-site mean-field approximations, to construct the
phase diagram of the BPCA for $p_3=1$ and $p_3=0$. The method 
detects both first- and second-order phase transitions, and also
can be used to locate DS transitions. The exponent values indicate that all
continuous phase transitions belong to the directed percolation universality 
class, while the exponent at the discontinuous phase transition agrees with 
the compact directed percolation value. 

We find evidence of a line of tricritical points in the 
($p_1,~p_2,~p_3$) parameter space, using the mean-field pair approximations. 
We also find a reentrant chaotic transition for $p_3=0$ in the mean field approximation,
that was confirmed by simulations. These observations illustrate the rich 
and at times surprising phase space structure found in simple nonequilibrium systems.

\noindent {\bf Acknowledgements}
We thank the Brazilian agencies CNPq and Fapemig for financial support of 
this work. 


\newpage

\begin{table}[htb] 
\caption{BPCA transition probabilities} 
\label{table:1} 
\newcommand{\m}{\hphantom{$-$}} 
\newcommand{\cc}[1]{\multicolumn{1}{c}{#1}} 
\renewcommand{\tabcolsep}{0.9pc} 
\renewcommand{\arraystretch}{1.2} 
\begin{tabular}{@{}lllll} 
\hline 
\cc{$\sigma_i / (\sigma'_{i-1}, \sigma'_i, \sigma'_{i+1})$}       
& \cc{(1,1,1)} 	& \cc{(1,1,0), (1,0,1) or (0,1,1)}	& \cc{(0,0,1), (1,0,0) or (0,1,0)}	& \cc{(0,0,0)}\\ 
\hline 
\cc{1}               & \cc{$p_3$} & \cc{$p_2$} & \cc{$p_1$} & \cc{0} \\ 
\cc{0}               & \cc{$1-p_3$} & \cc{$1-p_2$} & \cc{$1-p_1$} & \cc{1} \\ 
\hline 
\end{tabular} 
\end{table} 

\begin{table}[htb] 
\caption{Joint transition probabilities for two BPCA subjected to the same noise.} 
\label{table:2} 
\newcommand{\m}{\hphantom{$-$}} 
\newcommand{\cc}[1]{\multicolumn{1}{c}{#1}} 
\renewcommand{\tabcolsep}{1.5 pc} 
\renewcommand{\arraystretch}{1.2} 
\begin{tabular}{@{}lllll} 
\hline 
\cc{$\sigma_i$, $\tau_i$}       & \cc{(1,1,1;1,1,1)} & \cc{(1,1,1;1,1,0)} & \cc{(1,1,1;1,0,0)} \\
\hline 
\cc{1,1}          & \cc{$p_3$}  & \cc{$\min{(p_2, ~p_3)}$} 	& \cc{$\min{(p_1,p_3)}$}  	   \\ 
\cc{1,0}          & \cc{0}      & \cc{ $b=\max{(p_2-p_3,~0)}$}	& \cc{ $b'=\max{(p_3-p_1,~0)}$} \\ 
\cc{0,1}          & \cc{0}      & \cc{ $c=\max{(p_3-p_2,~0)}$}	& \cc{ $c'=\max{(p_1-p_3,~0)}$}  \\  
\cc{0,0}          & \cc{$1-p_3$}& \cc{1-$\max{(p_2, p_3)}$}	& \cc{1-$\max{(p_1,p_3)}$}	\\ 
\hline 
\end{tabular} 
\begin{tabular}{@{}lllll} 
\hline 
\cc{$\sigma_i$, $\tau_i$} & \cc{1,1,1;0,0,0} & \cc{(1,1,0;1,1,0)} & \cc{(1,1,0;1,0,0)}  \\ 
\hline 
\cc{1,1}         & \cc{0}      	&\cc{$p_2$}	    & \cc{$\min{(p_1,p_2)}$}     	 \\ 
\cc{1,0}         & \cc{$p_3$}   	&\cc{ 0}        & \cc{$b''=\max{(p_1-p_2,~0)}$} 	 \\ 
\cc{0,1}         & \cc{0} 		&\cc{ 0}        & \cc{$c''=\max{(p_2-p_1,~0)}$}    \\  
\cc{0,0}        & \cc{$1-p_3$}  	&\cc{$1-p_2$}   & \cc{$1-\max{(p_1,~p_2)}$}	 \\ 
\hline 
\end{tabular} \begin{tabular}{@{}lllll} 
\hline 
\cc{$\sigma_i$, $\tau_i$}	& \cc{(1,1,0;0,0,0)} & \cc{(1,0,0;1,0,0)} & \cc{(1,0,0;0,0,0)} & \cc{(0,0,0;0,0,0)}\\ 
\hline 
\cc{1,1}         & \cc{0} 		& \cc{$p_1$}  & \cc{0}     &\cc{0} \\ 
\cc{1,0}         & \cc{$p_2$}	& \cc{0}      & \cc{$p_1$} &\cc{0} \\ 
\cc{0,1}         & \cc{0}  		& \cc{0}      & \cc{0}     &\cc{0} \\  
\cc{0,0}         &\cc{1-$p_2$} 	& \cc{$1-p_1$}&\cc{1-$p_1$}&\cc{1} \\ 
\hline 
\end{tabular}
\end{table} 

\newpage

{\large \bf Figure Captions}

\noindent {\bf Fig. 1.}
Phase diagram for the BPCA, $p_3 = 1$. One- and two-site
mean-field approximations are compared with simulation results.

\noindent{\bf Fig. 2.}
Phase diagram for the BPCA, $p_3 = 0$. The mean field approximation predicts a 
tricritical point at ($p_2=2/3, ~p_1=1/3$),
where a discontinuous boundary (dashed line) meets a continuous 
transition line. 

\noindent{\bf Fig. 3.}
Line of tricritical points and the 
critical surface in the ($p_1,~p_2,~p_3$) parameter space, 
as predicted by pair level mean field approximation. The 
tricritical line ends near $p_3 \simeq 1/3$, while at the site level it 
extends to $p_3=0$, as shown in Fig. 2.
 
\noindent{\bf Fig. 4.}
Density of active sites and Hamming distance, in the mean-field approximation, 
at site level, for $p_3\!=\!0$. Note that for $p_2 > 2/3$, the transition is 
discontinuous.

\noindent{\bf Fig. 5.}
DS transition line, for $p_3\!=\!0$. The one-site mean-field approximation
(inset) is compared with simulation data (main graph). Reentrant behavior is 
observed in both cases. 

\noindent{\bf Fig. 6.}
DS transition line for $p_3\!=\!1$. The DS boundary meets 
the absorbing transition lines for ($p_1,~p_2$) near 1. 

\noindent{\bf Fig. 7.}
DS growth exponent for two different initial damages, 
for $p_3\!=\!1$. The left maximum corresponds to the absorbing transition
and the right maximum corresponds to the DS transition. 

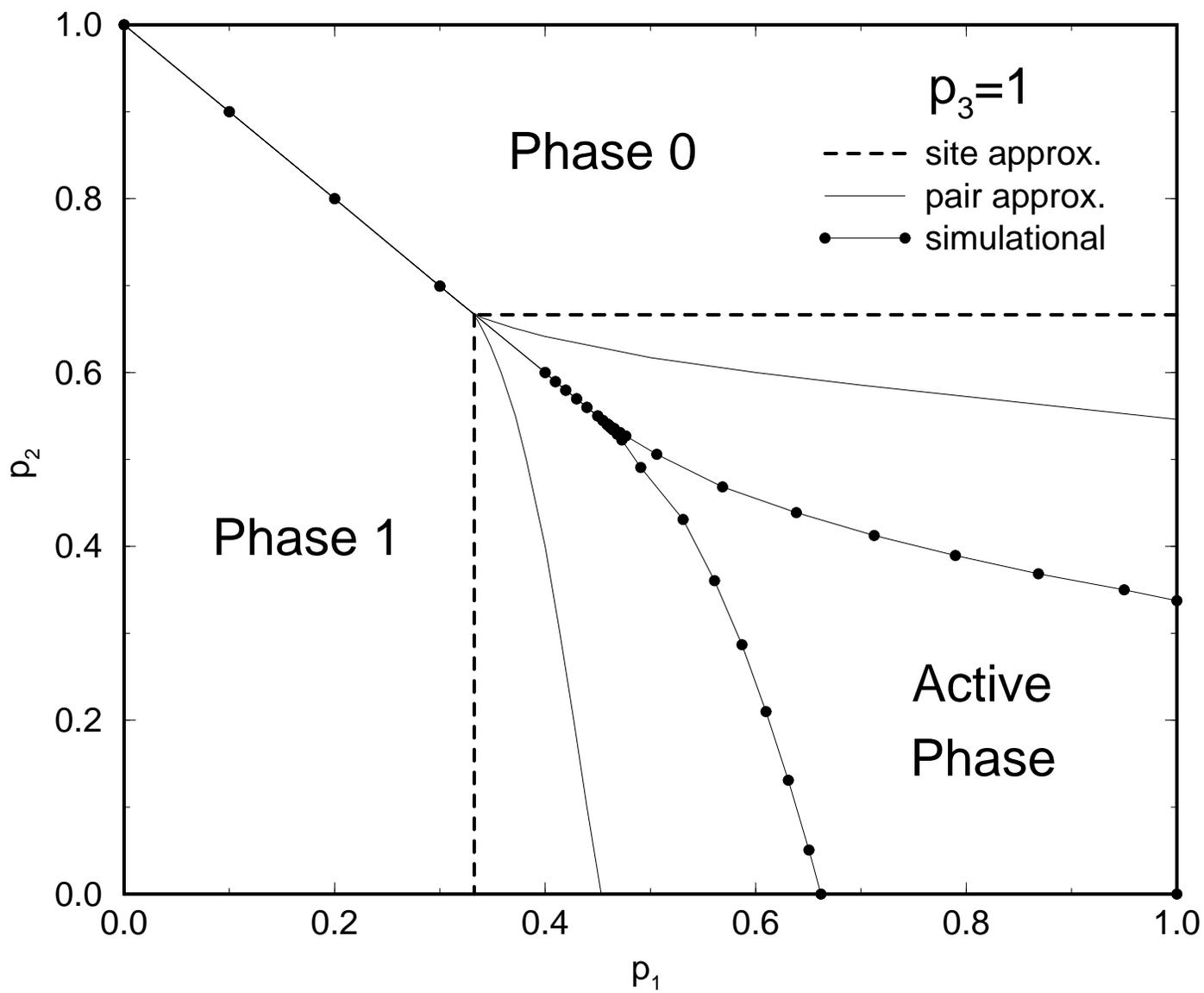
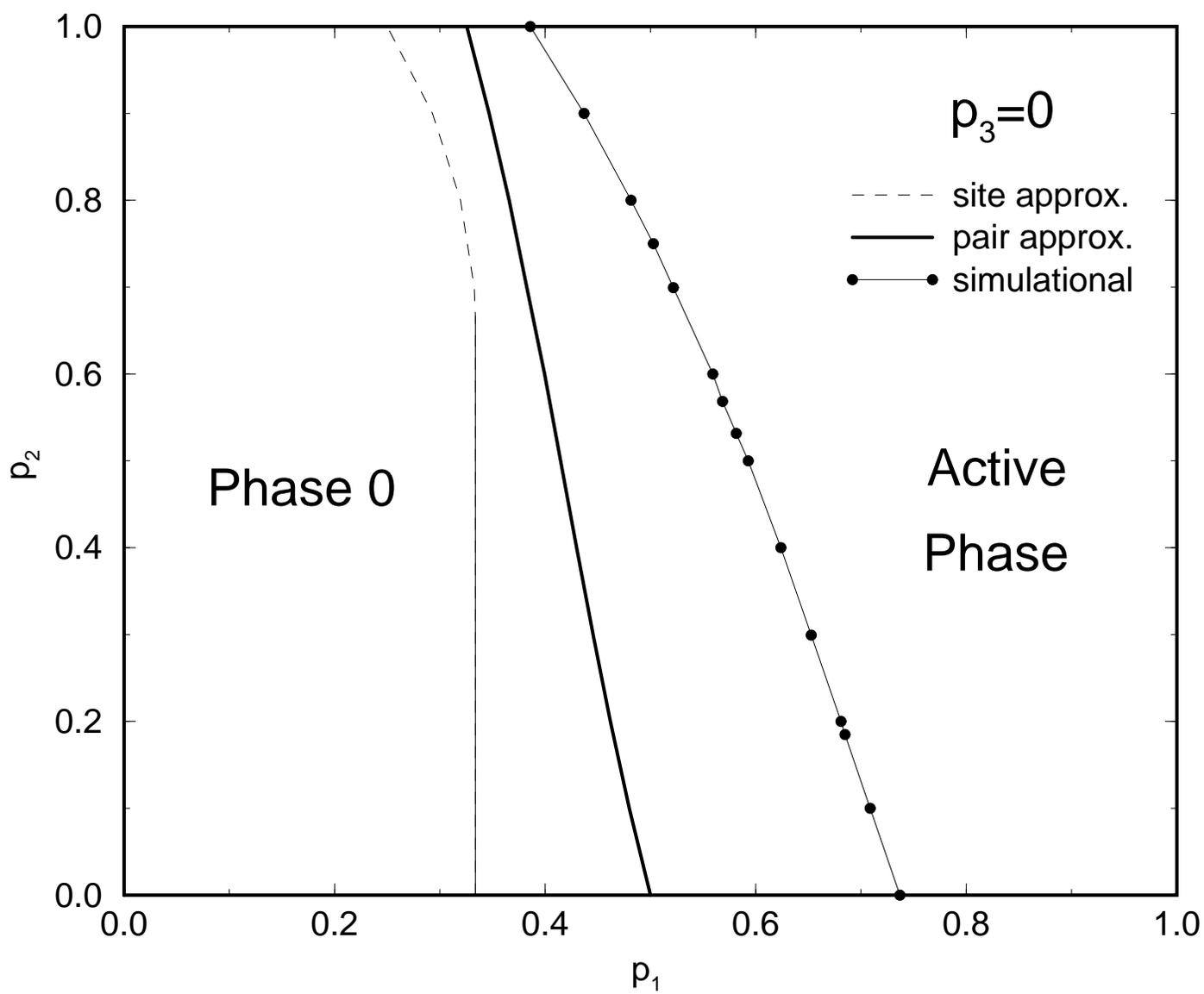
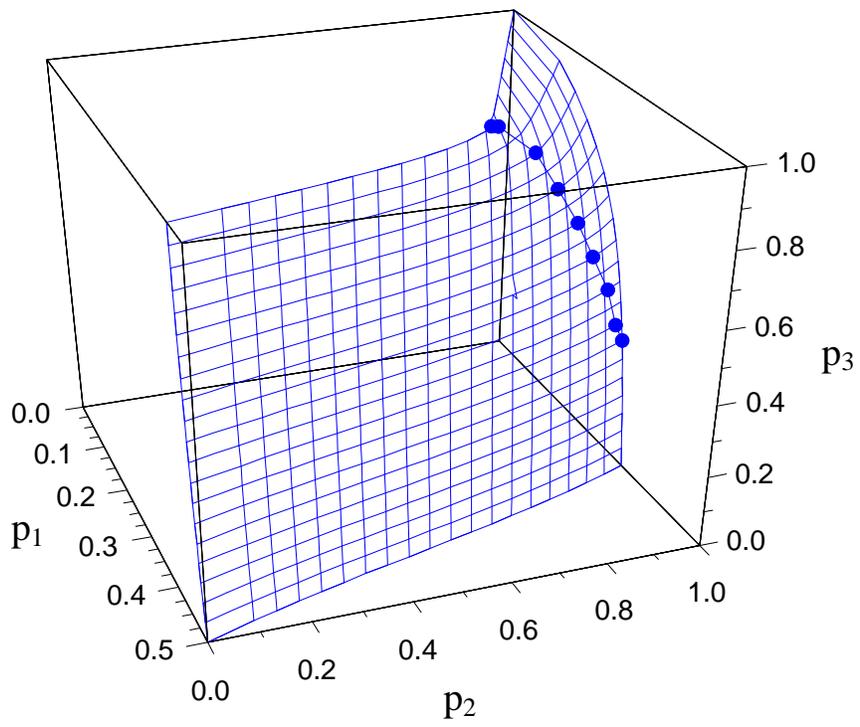
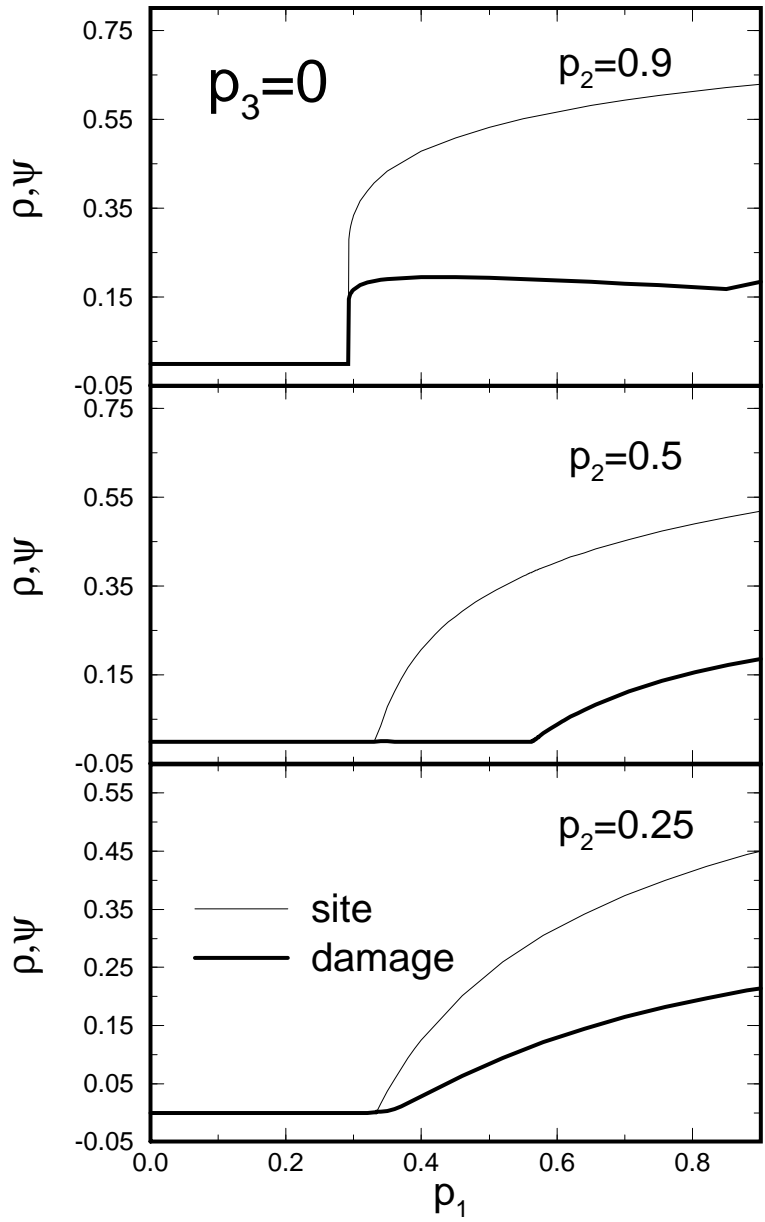
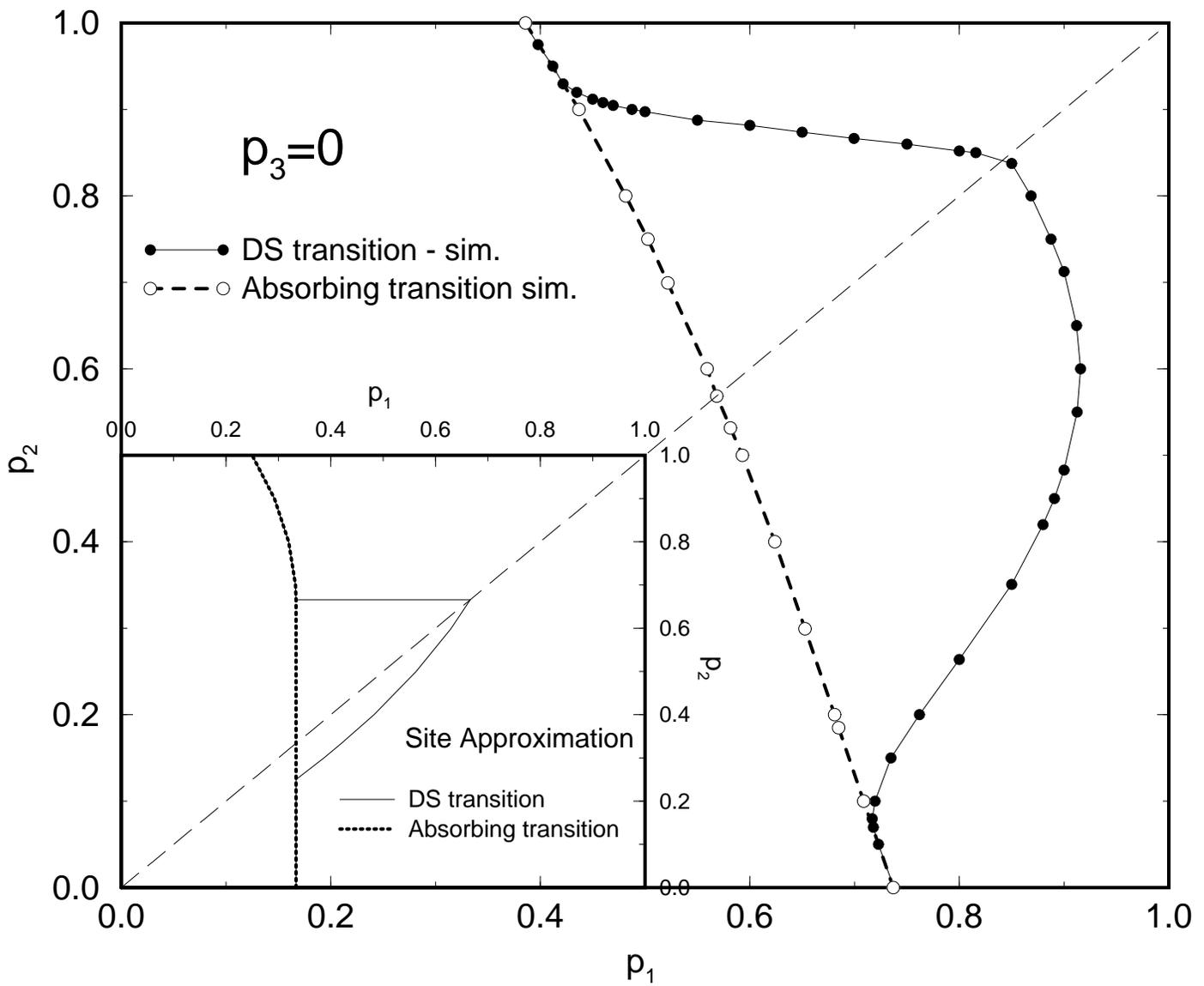
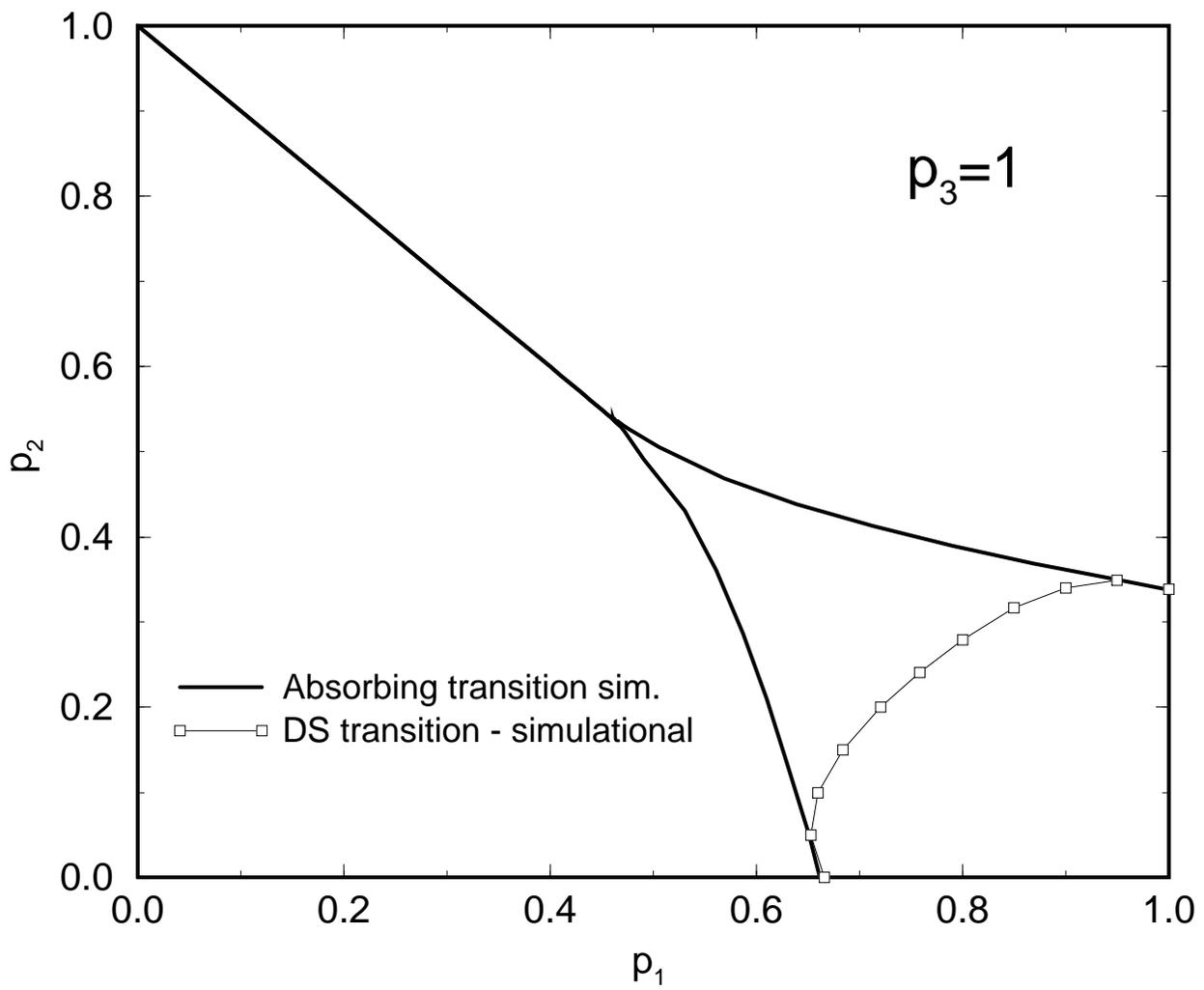
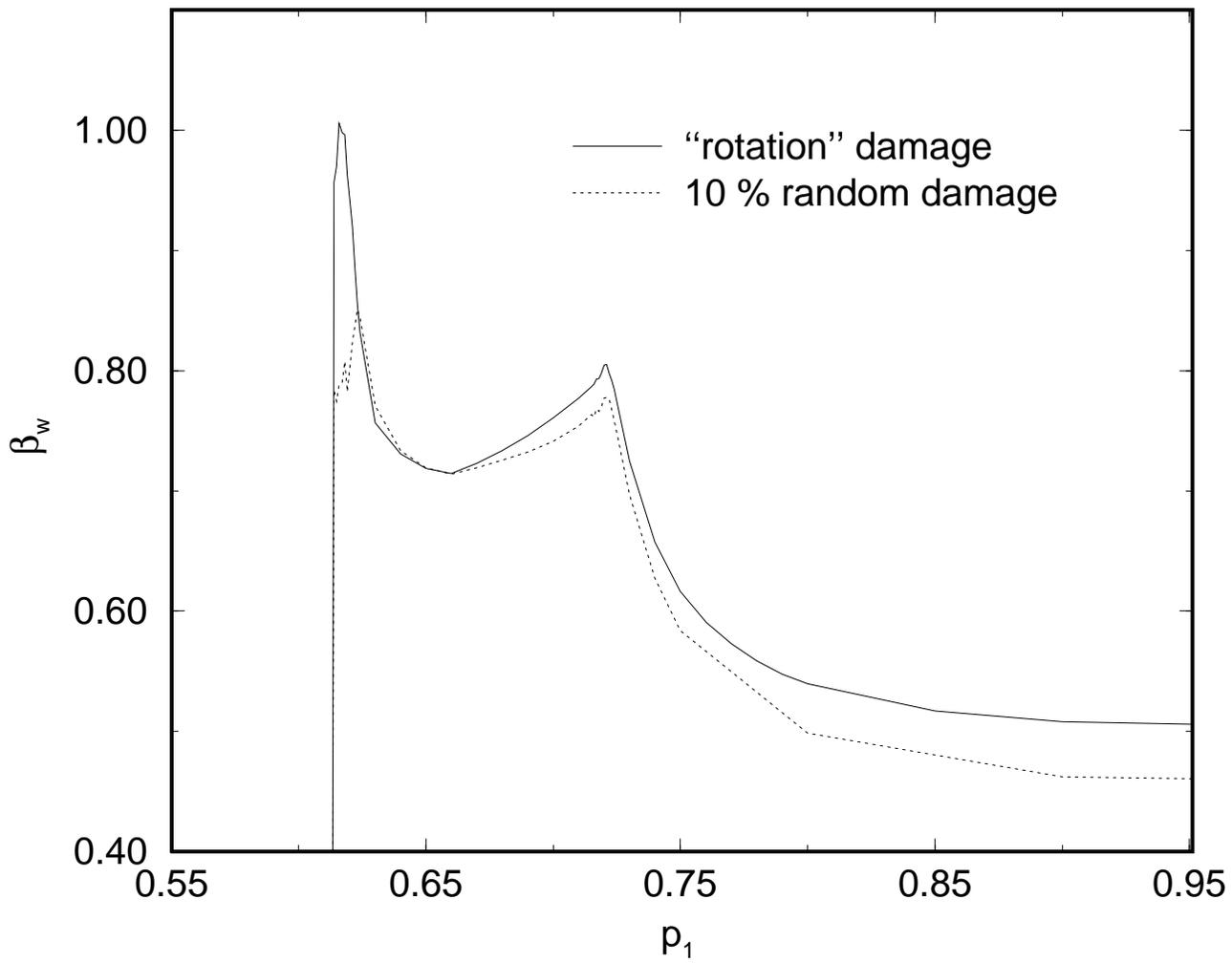
\end{document}